\documentclass{article}

\usepackage{latexsym}

\begin{document}

\centerline{\Large{A local interpretation of Quantum Mechanics}}

\vskip 0.5 truecm

\centerline{Carlos Lopez}

\vskip 0.3 truecm

\centerline{Dept. of Physics and Mathematics, Facultad de Ciencias}
\centerline{ UAH, Alcal\'a de Henares  E-28871 (Madrid, SPAIN)}
\centerline{carlos.lopez@uah.es}

\vskip 0.3 truecm

\begin{abstract}
A local interpretation of quantum mechanics is presented. Its main ingredients are: first, a label attached to one of the ``virtual'' paths in the path integral formalism, determining the output for measurement of position or momentum; second, a mathematical model for spin states, equivalent to the path integral formalism for point particles in space time, with the corresponding label. The mathematical machinery of orthodox quantum mechanics is maintained, in particular amplitudes of probability and Born's rule; therefore, Bell's type inequalities theorems do not apply. It is shown that statistical correlations for pairs of particles with entangled spins have a description completely equivalent to the two slit experiment, that is, interference (wave like behaviour) instead of non locality gives account of the process. The interpretation is grounded in the experimental evidence of a point like character of electrons, and in the hypothetical existence of a wave like, the de Broglie, companion system. A correspondence between the extended Hilbert spaces of hidden physical states and the orthodox quantum mechanical Hilbert space shows the mathematical equivalence of both theories. Paradoxical behaviour with respect to the action reaction principle is analysed, and an experimental set up, modified two slit experiment, proposed to look for the companion system. 

\vskip 0.5 truecm

\end{abstract}

\vskip 0.5 truecm

\section{Introduction}
%\label{intro}
Quantum Mechanics (QM), including Quantum Field Theory (QFT), is the most successful mathematical  framework of physical theories, with regard to its broad scope of applications and accuracy of predictions; but it is also a battle field of deep metaphysical debates. Besides the orthodox interpretation \cite{Born}, \cite{Heis}, \cite{vonNew}, \cite{Path},  different alternative schools have appeared \cite{Ein-Fra}, \cite{Bohm}, \cite{Everett}, \cite{Bunge}, \cite{Hartle}, \cite{Ball2}, \cite{espagnat}, \cite{GellMann}, \cite{Wheeler}, \cite{Ballent}, \cite{Home}, \cite{Gibbins}, \cite{Bellbook}, \cite{Penrose}, \cite{Omnes}, \cite{stoch}; to get a wide view of the present state of the art is a though undertaking (see for example \cite{Belifante}, \cite{WZ}, \cite{Ball3}, \cite{bell2}, \cite{Home-Whitaker}, \cite{biblioguide}, obviously not up to date). 

Alternative {\it theories}, i.e., with experimentally distinguishable predictions from QM, either have already been  ruled out or are out of scope of the present technology. However, the search of alternative {\it interpretations} of QM has not been left to academics in metaphysics. Why? Probably because the odd behaviours we find in QM are far away from the classical world, and our nearest understanding of nature seems to be denied by QM.
We can elaborate a list of paradoxes in QM, in order of relevance with regard to their challenge against classical and common sense concepts and knowledge. This list is obviously a matter of personal taste. Mine is

\begin{enumerate}

\item Measurement problem and the projection rule.

Measurement and the projection of state, as a physical phenomenon {\bf different} from other quantum interactions, is the source of many interpretative problems: subjectivity,
non local projection of state process for entangled systems, quantum classical boundary at the measurement apparatus, unknown definition of macroscopic system, etc.

\item Non local interaction.

Special (SR) is as firmly established as QM. However, we should agree that some natural law breaks relativistic invariance if and only if there was found unambiguous  evidence of a physical non local interaction. Bell's type inequalities \cite{Bell}, \cite{Clau}, as well as inconsistencies in assignment of pre--measurement (hidden) values to some families of operators \cite{GHSZ}, \cite{mermin} (GHZ theorems), are considered indirect evidence of the spooky action at distance. 

\item Wave particle duality

 Fact is we observe particles (spot in a screen or photographic plate, path in a cloud chamber) in individual measurements, and diffraction patterns, as in the two slit experiment, exclusively in statistical samples.

\item Tunnel effect.

Again,  a broken fundamental law, conservation of energy. QM does not describe the evolution process of the system between both sides of the potential well,  just probabilities. Initial and final energies match; however, the hypothetical process is not understood, energy fluctuations is perhaps the best description we can find.

\end{enumerate}

Since the beginning of  QM formulation in the nineteen twenties these and other questions opened a debate about the very meaning of scientific knowledge: What a scientific theory is?  What an interpretation provides?
What elements of reality are represented in the mathematical formulation?  What do we do of unavoidable mathematical ``artifacts'' without obvious physical meaning? The EPR program \cite{EPR} is a  generic proposal about these subjects; two explicit conditions where listed, reality and completeness. Another, locality, was implicit in the arguments (``\ldots without in any way disturbing a system''). Locality can be encircled in a more generic requirement of conservation laws, which should obviously be rejected if the corresponding symmetries were violated. Reality and completeness are more conceptual, metaphysical, they represent   basic requirements we would ask to a scientific theory \cite{EPR}:

\begin{enumerate}

\item Completeness: {\it Every element of the physical reality must have counterpart in the physical theory}.

\item Reality: {\it If, without in any way disturbing a system, we can predict with certainty the value of a physical quantity, then there exists an element of physical reality corresponding to this physical quantity}. 

\end{enumerate}

The {\it gedanken} experiment presented in EPR, where position and momentum of an isolated particle {\bf could} be predicted with certainty through measurements in its entangled pair, was argued as evidence of the incompleteness of QM, where both magnitudes do not commute and can not have simultaneously precise values. Advocates of the orthodox interpretation stood behind the completeness of QM. First, the ``{\it could}'' was interpreted as these magnitudes not actually having values {\bf before} the arbitrarily chosen measurement on the other particle. Second, it was increasingly apparent that QM (at least in its orthodox interpretation) is not a local theory, an implicit ingredient in the EPR argument. The particle physical state is perturbed, in a non local way, by its companion measurement.

Completeness, reality, and locality are the EPR program for an interpretation of QM. However, another ``implicit'' condition seems to have been amply imposed by researchers on the subject, existence (or not) of a classical probability distribution associated to hidden variables. There is not any reference to this condition in EPR, the program is open to any kind of theories, as far as they fulfil the three requirements.

Along the way there have appeared  evidences against hidden variables theories. We can group them in two main groups, the probabilistic arguments \cite{Bell}, and the inconsistencies when assigning hidden values to some families of ``physical'' magnitudes (self adjoint operators) \cite{GHSZ}, \cite{mermin}. Bell's type inequalities theorems show the mathematical impossibility of finding classical probability distributions for hypothetical local hidden variables matching the QM (and experimental \cite{Aspect}) correlations. Theories whose formulation is not probabilistic in the classical sense  are not banned; QM is one of them, when using these weird objects named amplitudes of probability. GHZ and similar theorems apply to families of non local self adjoint operators. If the EPR program is to be completely fulfilled, including locality, these non local objections will be inconsistent. How can exist a pure quantum state eigenvector of non local operators if the initial conditions were reached through local interactions and the evolution is also local (a kind of superselection rule)? In an interpretation of QM fulfilling the EPR program, non local self adjoint operators can not represent elementary physical magnitudes; 
only local magnitudes (and not necessarily all self adjoint operators, there is not explicit prescription about this point in QM) can be measured in local measurements (interactions). 

Subjective interpretations, even if they are not popular, make very re\-le\-vant the task of finding an interpretation of measurement as a regular interaction, not different from others, and to restore objectivity in Physics, at least as a plausible hypothesis of work. Measurement presents difficulties not only when the projection rule must be applied, where the output state is different from the input; in the simplest measurement of a physical magnitude in an eigenstate of the corresponding operator we find a paradox, there is a change of state in the measurement system (the pointer) while, according to QM, input and output states of the measured system coincide. The action reaction principle seems to be violated. Also in indirect measurements, when some ``virtual'' path of the system is obstructed; even when, in an individual event, there is not interaction with the additional system (obstacle or measurement apparatus), the system under study changes of state, for example with a behaviour that was forbidden in absence of the obstruction. Action reaction principle should be added to locality as possibly violated laws in the orthodox interpretation of QM.

The paper is organised as follows: next section quickly reviews some experimental evidences in quantum mechanics. Then,  a list of paradoxes (measurement and projection rule, entanglement and non locality, two slit experiment and action reaction principle, and wave particle duality) is discussed. A mathematical formulation of quantum mechanics with hidden variables, grounded in the path integral formalism, is proposed in Section 4; both 
enlarged Hilbert spaces for the spinless point particle and for the spin states are introduced, and Born's rule for relative frequencies applied. It is explicitly shown the parallelism between entangled spins and the two slit experiment under the proposed interpretation; i.e., the hypothetical non local behaviour of quantum mechanics becomes in this interpretation a local interference phenomenon. An experimental set up to detect the mysterious companion is proposed, inspired in the action reaction principle.

%%%%%%%%%SECCION 2%%%%%%%%%%%%%

\section{Experimental evidences and metaphysical debates}
%\label{exp}

Let us review some evidences. One, electrons are point particles. This is scientific, inductive evidence. Of course it must be understood as approximate, with finite accuracy, etc. Individual electrons (and photons) imprint a point like spot on a screen or photographic plaque, and a unidimensional path on a cloud chamber; we sometimes find wavelike patterns made of individual spots. QM is so puzzling that we are uncertain of applying the scientific method of induction here. Each time the mass of an electron is measured similar values are  found. Of course direct evidence is just about these individual electrons actually measured, but each new measurement increases (by induction) our confidence in the existence of a universal constant $m_e$. It would be enough {\bf one} different measured value (and I am sure it would be analysed and repeated with the most extreme care) to reject this hypothesis. Scientific rules of the game are this severe, and that is why scientific knowledge is so robust. A wavelike (or any other not particle like) imprint of an individual electron will reject the particle like hypothesis. Meanwhile, electrons are point particles. 

Of course, electrons are point particles when we observe them. This is almost a tautology, but again QM paradoxes have made us insanely careful. When we use detectors in the two slit experiment to observe the electron in the slits, it {\bf always} goes through one or the other slit, never through both or any other way. This is consistent with the previous evidence of electrons as point particles. Yet, we avoid to assert that this is also the case in absence of detectors. In QM there is not detailed description of the process. 

The two slit experiment is also scientific evidence of a wave like behaviour of electrons.  These scientific facts, which seem contradictory, are a challenge to our ability to build models, but to deny raw facts (spot), or to state that things are different when we do not look
(a strongly subjective statement)  does not seem a satisfactory explanation. The EPR program has not expired. The point particle 
property is an element of reality, and to incorporate it into QM, even if all its predictions are preserved (that is, just as a metaphysical addendum) has scientific interest and could be source of inspiration for new theories.

Even if scientifically unsuccessful,  the (metaphysical) debate is sometimes valuable because it fosters a deep analysis of many interesting points that usually do not deserve our attention, but become instructive after a more observant study. As an example, consider the concept of (or alternative approaches to)  probability \cite{propensity}: either as  relative frequencies in a finite universe of different elements, or as  ``propensities'', probabilities that identical elements of a considered universe entail to show different properties when observed.

Imagine a large black box with $2N$ balls; you are informed that either (1) $N$ are red and $N$ green, or (2) each one is white and has equal probability (propensity) of becoming red or green {\bf as soon as} it is observed. Extract balls one at a time. If you get $N+1$ red or green balls, the first case is discarded. You can end up with $N$ red and $N$ green balls in both alternatives, but this is very improbable in the second case for large $N$, so you could guess (1). There is a practical impossibility to perform the former method for very large $N$, say of the order of  $10^{23}$; with a limited sample, just a small fraction of the total set, there is no way to distinguish (1) from (2). 

 In EPR experiments (e.g., entangled spins of two particles) it seems  the propensity concept wins, because there are no {\bf local} hidden variables with a classical distribution of relative frequencies able to match the QM and experimental correlations. After Alice measurement is performed there is some  transmission of propensities to the second particle through the projection rule, a non local process. However, it could also be  transmitted, again in a non local way,  the future output of  Bob's measurements in arbitrary directions, following a table of relative frequencies in a draw. EPR experiment is about local versus non local behaviour (and completeness of QM, and elements of reality), not about propensities (represented in the quantum state) versus relative frequencies (represented through, e.g., non local hidden variables). No experimental evidence in favour of propensities exists (can exist). In fact, QM is formulated in term of amplitudes of probability, not probabilities (propensities or relative frequencies), and the usual interference phenomenon can not be reproduced in classical probabilistic formulations. 

In summary, we have inductive evidence of the point like character of the electron, and of a wave like behaviour in some experiments, which seems contradictory. We do not have evidence about the propensity versus relative frequency rival interpretations. The de Broglie Bhom model \cite{Bohm} shows that hidden variables theories can not be rejected because of experimental predictions
(complete predictive agreement), and the point like behaviour of the electron pushes us to consider either hidden variables or a subjective theory (electrons are point particles {\bf when} observed).  
Bell's type inequalities teach us an important lesson: the electron physical states with hidden spin variables (prescribed output for {\bf all} measurements) can not be a {\bf complete} description of the system, i.e., a set of independent, isolated, states. But hidden spin state and interference, so amplitudes of probability, is not ruled out.

%%%%%%%%%SECCION 3%%%%%%%%%%%%% 

\section{Paradoxes in Quantum Mechanics}
%\label{paradoxes}

\subsection{Step interaction, and measurement}
%\label{step}

We want to interpret measurement as any other interaction in QM, with the obvious property that the measurement apparatus modifies its pre--mea\-su\-re\-ment state into a macroscopically different post--mea\-su\-re\-ment state  (the pointer), independently of the presence of an observer.  If the projection rule should be applied to all interactions of small quantum systems with macroscopic systems (precisely defined), with explicit and objective quantum rules, the problem of measurement would not exist. But this is not the orthodox interpretation of QM. The projection rule represents either a subjective process associated to the observer, or, in some alternatives, a physical phenomenon external to QM, where decoherence, thermodynamics or some unknown boundary between CM and QM, explains the change of quantum state as a short walk outside QM, the system returning to the quantum world in a state different from its departure, giving way to the projection of state, non unitary jump.

It is paradoxical that CM can be both an approximation of QM and an independent physical theory (at least when applied to quantum measurement with macroscopic systems). It is paradoxical that Statistical Mathematics can  give way to some fundamental physical law, outside QM, violating the QM dynamics. Can an ensemble violate the fundamental interaction and evolution rules of its simple constituents?

\subsubsection{Classical Hamiltonian step interaction}

Let us consider a step interaction in Classical Mechanics, along a short time interval $\tau$. In Hamiltonian formulation (the Lagrangian approach is equivalent; in Newtonian Mechanics we should take into account the action reaction principle) and for small $\tau$, a simple model is the interaction Hamiltonian $H_{int} = \lambda q y$, where $q$ and $y$ are variables of both systems in interaction respectively, and $\lambda$ represents an intensity of interaction; equivalently, $\lambda = 0$ for $|t|> \tau /2$. Hamiltonians $H_1$ and $H_2$ determine the dynamics of both isolated systems, but we will not consider the additional ``free'' evolution along $\tau$ associated to them, neither do we consider it to be negligible, although in a step interaction it usually is.  Our interest is focused on the additional evolution associated to the interaction,  more precisely on its first order approximation. 

There is a step evolution for all variables $p_q$ and $\pi _y$ of each system which do not commute with $q$ and $y$, 
$\bigtriangleup p _q   = \{p_q, H_{int}\} \tau =  \{ p_q, q\} (\lambda \tau )y$, and
$\bigtriangleup \pi _y   = \{\pi _y, H_{int}\} \tau = \{ \pi _y, y\} (\lambda \tau )q$. As we see, this step is proportional to the interacting variable of the other system, so it can be used to distinguish among states of each system with different $q$ and $y$ values. $q$ and $y$ are constant (up to their isolated systems dynamics) along $\tau$, but for $H_1(p_q)$ and $H_2(\pi _y)$ there will be  an evolution of $q$ and $y$
after the interaction different from the isolated system evolution, because of the  increments $\bigtriangleup p_q$ and $\bigtriangleup \pi _y$. If we consider the second system as a measurement apparatus, it is designed to transform  $\bigtriangleup \pi _y$, or $\bigtriangleup y$, into a new macroscopic state of the pointer. It is unavoidable that the output state of the measured system will be different from the input. It is a technological challenge to minimise this perturbation, $\lambda \tau \to 0$, and yet get a macroscopic response. There is, however, no fundamental limit of accuracy in CM. 

Of course, $H_{int}$ is an idealisation, and real measurements are something like $H_{int} + H_c$, with $H_c$ some additional term depending on the specific experimental set up. We can apply a step interaction on an ensemble of systems $1$, and  classify them into sub--ensembles according to the output. It is clear that the specific protocol we choose to do it will in general give way to different  sub--ensembles, even though the ``relevant'' term is an ideal $H_{int}$. The probably unavoidable $H_c$ introduces different outputs for different techniques, no matter if the measured magnitude is $q$ in all of them. 

\subsubsection{Quantum step interaction}

In QM we face  a quite different phenomenon. First, there is a minimal (inter)action, $H \tau \geq h$, Planck's constant; this property makes explicit the discrete and discontinuous character of many QM interactions. Second,  perhaps related to the former, ideal measurements can be performed; when a physical magnitude is measured, the final quantum state is common to different measurement techniques, at least for discrete magnitudes. So, there is something fundamental in these interactions, independent of particular details. If there is a discrete, discontinuous step, it seems reasonable that small differences between measurement processes for the same magnitude do not (can not) give way to forbidden small differences in the output. Quantum interaction (and therefore measurement) becomes quite different from its classical counterpart at this small, quantum, scale; at larger scales we can recover the classical approximation when the discrete, discontinuous, phenomenon can be approached by the continuum. Magnitudes as spin have, on the other hand, no classical counterpart. 

The problem with the projection rule under measurement is not about perturbation of the measured system, this perturbation is also a classical phenomenon; neither it is about discontinuities without classical counterpart, some quantum discontinuous steps could be represented by unitary maps. The problem is that measurement is not like other interactions, it is not a unitary map. Probably, the best prescription at hand for the projection rule is that {\it everybody knows when a measurement happens}. 

Hidden variables theories state that measurement is like any other interaction, that a hidden difference between individual systems in the same pure quantum state (so, an ensemble) determines the difference in the outputs. No role of observer, no unknown boundary between QM and other  worlds. Although there are well known difficulties with hidden variables theories and their interpretation, it should be acknowledged its success in this particular point. 

 Some people reject hidden variables  because they can not give a local description of entangled spins correlations, according to Bell's inequalities. But the measurement problem has enough entity by itself to ponder alternative interpretations, even if they do not solve all other paradoxes in QM, as non locality. We can reject hidden variables theories because of Ockham's razor or whatever other reason, not because their  reach is limited to some, and not all, paradoxes.

There is an argument in favour of hidden variables  I have not found in the literature.  It does not refer to the projection rule when the input system is not  an eigenstate of the measured magnitude (the difficult challenge of violated  unitary evolution under measurement), but to the simples measurement of a  pure quantum  eigenstate. The output quantum system equals the input, according to QM rules. It is something impossible at the classical level, there is necessarily some change of state. Hidden variables could change at each individual measurement, in such a way that the overall description of the ensemble, the quantum eigenstate, was maintained. Hypothetical perturbed hidden variables, not commuting with the measured magnitude, do not have (in QM) definite values  before and after the interaction; if they had a hidden value and it was modified by the measurement, it would remain hidden. 
Does  not apply the action reaction principle to this process in QM? The pointer state is modified after measurement because of the interaction, not as an isolated evolution of the apparatus. Where is the reaction, unavoidably associated to an interaction, in the measured system?

\subsection{Entangled spins, and non local interaction}
%\label{spin}

Let us analyse now the experiment (we will denote it EPR) of generation, at event O, of two spin $1/2$ entangled particles $a$ and $b$ in a null total spin state, followed by measurement  of spins at spatially separated events A and B. I  analyse  three different interpretations,  not to conclude that one of them is preferred but because there are alternatives and perhaps they tells us  something about which to ponder.  A preliminary version of the content of this section was presented in \cite{S3}.

\begin{enumerate}

\item  First, the orthodox interpretation. A statistical ensemble of measurement e\-vents A and their outputs determine projection of state of the compound system and, as a consequence, a modified table of probabilities for an ensemble of measurements in B, or {\it vice versa}. If the projection rule is a real physical process the phenomenon is non local. It is difficult to state an {\bf individual} change of state in particle $b$ when this state is characterised by a table of probabilities, we necessarily need a statistical sample to detect the change of state. Does change each individual state of particle $b$ after each individual  measurement of particle $a$, or are we selecting a conditional subensemble in the statistical sample? The heart of EPR is about this dilemma. If we could state unambiguously an individual change of state in particle $b$ there would be no doubt about non locality. But we can not. Even when a physical state is described by a table of probabilities, it is possible to detect a physical change of state of an individual system in two ways:

\begin{enumerate}

\item Some output that was forbidden in the initial state becomes possible (not necessarily certain) later on. An individual process in which this previously forbidden output occurs shows unambiguously a change of state. We do not need statistics here.

\item Some output that was certain becomes uncertain after the process. Each individual process in  which the previously certain output does not occur also shows  unambiguously a change of state.

\end{enumerate}

None of these ways can be applied in EPR experiment. There are no, previous to the measurement event A, forbidden or certain outputs for spin measurement at  B. A change in the table of probabilities can not be, up to the previous cases, stated unambiguously in  a single measurement event. Notice that certainty, probability null or unit, for some output of measurement in B after (better once known) measurement output in A, is not conclusive about a change of state, as far as this output was already possible previously to A measurement.

\item  Local hidden variables are ruled out because of Bell's type inequalities. There are not local hidden variables with classical probability distributions able to match the QM and experimental correlations. This is mathematical fact, you can not beat a theorem. 
However, it is quite easy to solve the linear equations for a table of hypothetical quasi probabilities reproducing QM correlations.
In the simplest non trivial case of three planar directions of spin, $\theta _j, j=1,2,3$, the solution is

\[
W(s_1,s_2,s_3) = \frac {1}{8} [ 
1 + s_1 s_2 \cos (\theta _2 - \theta _1) + 
\]

\[
+ s_1 s_3 \cos (\theta _3 - \theta _1) + 
s_2 s_3 \cos (\theta _3 - \theta _2) ]
\]
where $s_j=\pm$ represent the spin state in direction $\theta _j$. We understand that the state of  particle
$b$, $(-s_1, -s_2, -s_3)$,  is opposite to that of $a$ in a total null spin state. $W$, and not $P$, is used to represent weights (or Wigner), because they can be negative.  $W$ is a mathematical artifact that can not be measured. Observable relative frequencies are correctly obtained

\[
P(s_1,s_2) = W(s_1,s_2,+1) + W(s_1,s_2,-1) = \frac {1}{4} \left( 
1 + s_1 s_2 \cos (\theta _2 - \theta _1) \right)  
\]
so, $\frac {1}{2} \cos ^2((\theta _2 - \theta _1)/2)$ for $s_1 s_2 = +1$, or $\frac {1}{2} \sin ^2((\theta _2 - \theta _1)/2)$ for $s_1 s_2 = -1$. 
For $\theta _j = (j-1) \pi /3$, $j=1,2,3$, we get $W(+,-,+)=W(-,+,-) = - 1/16$ and $W=3/16$ otherwise. Negative weights are unavoidable.

It could be argued (but I do not stand behind it) that it is better to loose positivity of $W$ in order to preserve locality. $W$ represents the spin equivalent of Wigner's quasi probability distribution in phase space \cite{Wigner}, reproducing QM correlations with a quasi probability distribution function having both positive and negative values.

\item  We  can maintain the hidden variables description of states $(s_1,$ $s_2,$ $s_3)^a$,  $(-s_1,$ $-s_2,$ $-s_3)^b$, and instead of  a table of probabilities,  assign  them a table of amplitudes of probability. After all, this is the orthodox QM way,
but now with additional variables.  The solution is even simpler,

\[
\Psi (s_1, s_2, s_3) = \sum _{j=1}^3 s_j e^{i \theta _j}
\]
in a not normalised representation. As usual with amplitudes (Born's rule), 

\[
P(s_1,s_2) = \frac {|\Psi (s_1,s_2,+) + \Psi (s_1,s_2,-)|^2}{
\sum _{s'_1,s'_2=\pm}|\Psi (s'_1,s'_2,+) + \Psi (s'_1,s'_2,-)|^2}
\]
is the probability, relative frequency of the output. If we could measure three spin directions, the corresponding hypothetical probabilities would be 

\[
|\Psi (s_1,s_2,s_3)|^2
\] 
adequately normalised. It is interesting to notice that the negative weight $W(+,-,+)= -1/16$  for angles $\{0, \pi /3, 2\pi /3\}$ becomes $\Psi (+,-,+)=0$. We can not measure simultaneously two directions in particle $b$, but one of them is indirectly (and independently) measured in the companion particle $a$,
``without in any way disturbing the system'' $b$. 

What we have found is a description of EPR with hidden variables and amplitudes of probability as machinery to calculate relative frequencies of outputs. If we understand $(s_1,s_2,s_3)$ as fixed (but hidden) in an individual experimental event, the output of  measurement is fixed from the generation event $O$ of the entangled pair; so, no non local interaction between A and B measurement events. However, as we are calculating relative frequencies through the amplitudes of probability, and it is the ``interference'' term in $\Psi (s_1,s_2,+) + \Psi (s_1,s_2,-)$  that allows to reproduce the QM correlations, we should perhaps conclude that there is an interference interaction between {\bf different} generation events, i.e., $O_1$ with $O_2$, etc.

I am sure this interpretation will not be popular, but it is interesting that it exists. After all, while A and B are spatially separated if we want to check non locality (they can be at both far ends of the observable universe), $O_k$ are usually in a temporal, causal space time relation. But this is not the point, we could obtain the same result if, instead of using a unique experimental set up and repeated measurements with it, we prepare a large number of identical copies of the apparatus and perform just one experiment with each one; now these $O_k$ can also be  spatially separated. 

Another interesting point of this interpretation is that non locality can be generalised,  extended to other QM interactions and measurements, e.g., the two sit experiment. We usually interpret it as an interference phenomenon between both components (left and right slit) of the amplitude function, without any relationship with a non local phenomenon. We can split both components into different individual events (an individual electron arriving to the screen), and understand the diffraction pattern as an interference between different measurement events; after all  we know that, when we look, the electron always goes through one or the other slit, never through both. The two slit experiment can also be performed with $N$ copies of the experimental set up, spatially separated: we could say there is just one two slit experiment (after rescaling of relevant distances and momentum of the electrons) that is successively  performed every year at different universities and research centres, for example for pedagogical reasons. If we collect all data of two slit experiments to the present time (an rescale appropriately) there would  certainly be a much better sample size than usual. 

This line of reasoning can be put upside down. If other quantum interference phenomena can be read as non local like EPR, we could also read EPR as a local interference phenomenon. Each particle could have spin wavelike behaviour, as it has spatial wavelike behaviour. I do not propose waves of particle $a$ interfering with waves of particle $b$, a non local phenomenon;  waves of particle $a$ among themselves, and {\bf independently} waves of particle $b$ among themselves, each $a$ and $b$ having a copy of the family of amplitudes $\{ \Psi (s_1, s_2, s_3)\}$ together with the individual (labelled) particle spin states $(s_1, s_2, s_3)^a$, $(-s_1, -s_2, -s_3)^b$.

\end{enumerate}

Notice the equivalence with the two slit experiment. The electron going through one slit, and wave interference of both amplitudes. Hidden paths through left or right slit are in correspondence with the hidden spin value $s_3=\pm$ in a third direction. 
The probability distribution is 
$|\Psi _L + \Psi _R|^2$, and interference gives way to some regions of the screen being avoided by the electrons, although $|\Psi _L|^2$ and  $|\Psi _R|^2$ do not vanish there. We do not need Bell's type inequalities to conclude that no classical probability distribution exists with two {\bf independent} strictly positive probabilities adding to zero.

\subsection{Two slit experiment, and action--reaction principle}
%\label{twoslit}

In the path integral formalism \cite{Path} the sum of virtual paths from right slit into a particular region of the final screen determines $\Psi _R$ and the sum of paths from the left one to the same region determines $\Psi _L$. When there is not measurement in the slits determining which slit goes the electron through, the relative frequency of spots in the region of the screen is $|\Psi _R + \Psi _L|^2$. On the other hand, if we have detected either directly or indirectly the electron in the slits, the distribution of spots is $|\Psi _R|^2 + |\Psi _L|^2$. We can perform a measurement restricted to R slit, e.g., by using a light beam in such a way that all electrons going through R are detected (in an ideal approach) while no electron through L is detected. In case an electron arrives to the screen without being detected in the slit we conclude it has gone through L, and apply the projection rule even though there has been no direct detection; this is an indirect measurement, and in orthodox QM the projection rule applies to both direct and indirect measurements. Let us consider a small region of the screen where
$|\Psi _R + \Psi _L|^2 \simeq 0$, and a particular event in which, with the light beam switched on, a not detected electron   arrives to this region. This is one of the cases in which we can unambiguously conclude that  a change of state has happened. But there has been no observed reaction on the additional system, the light beam. Is the action reaction principle violated in this process? 

Similar experiments have been performed with spin. Four Stern--Gerlach systems are prepared, numbers 1 and 4 in $X$ direction
and numbers 2 and 3 in $Y$ and $-Y$ directions respectively. The electron beam out of 1 corresponding to $+$ spin in $X$ direction (the $-_X$ beam is discarded) goes then through 2 and 3, an finally through 4. If we do not 
block one of the beams (with $+$ and $-$ spin in $Y$ direction) between 2 and 3, and being both processes opposite,
the initial $+_X$  state is reconstructed (no projection rule) and final measurement after 4 gives always $+_X$ output. 

If we block one of the beams, say $-_Y$, electrons arriving to 3 are in $+_Y$ state, and measurement in 4 gives $+_X$ and $-_X$ with even probabilities. Let us concentrate in a single event. An electron does not hit the initial obstacle after 1, neither the intermediate obstacle (both obstacles able to detect the particle) between 2 and 3, and after 4 it hits the $-_X$ detector. An event that was forbidden in absence of the intermediate obstacle (which allows the indirect measurement) happens in its presence, so we can unambiguously state a change of state associated to the presence of the obstacle. Again, no observed reaction associated to the action on the electron.

There seems to be something odd with the action reaction principle in QM, at least for indirect measurements. In the two slit experiment the theories of hidden variables will point to L (respectively R) as the hidden variable in $\Psi _R + \Psi _L$ for an individual event; but we can not understand $\Psi _R$ or $\Psi _L$ separately as a complete description of the individual electron state. Even with L as hidden variable determining the path of the electron, we need both $\Psi _L$ and $\Psi _R$ as a statistical representation of something real, an accompanying wavelike system, because $|\Psi _L |^2 + |\Psi _R |^2 \neq |\Psi _L + \Psi _R|^2$. The reaction on the light beam for the indirect measurement in the former experimental set up could be exerted by this hidden companion. Taking into account that the diffraction pattern disappears, we could make the hypothesis that it is this wavelike companion which suffers a phase shift (action), so that the hypothetical spatial hidden diffraction pattern also shifts by a stochastic  distance. We only see one spot of each individual diffraction pattern. The shift is  stochastic, different at each individual event, and the overall diffraction pattern disappears in the statistical sample of measurements.

\paragraph{Proposed experiment.} If the action were a phase shift on the wavelike hidden companion,  the reaction could be a phase shift on the photon. A laser light beam of coherent photons could then be used to detect a phase shift (decoherence) whenever an undetected electron arrives to the screen. 

The question of what is the reaction on the light beam associated to the unambiguous action over the electron remains obscure. But facts are conclusive; without light beam, electrons do not hit on some forbidden areas of the screen; with the light beam switched on some undetected electrons do. Also detected ones, but we are focused on the ``hidden'' action reaction. This is not just a modified table of relative frequencies, it is an individual effect on an individual electron, whose  actual end point on the screen was ruled out in absence of the additional system. There is (inter)action.   It is the light beam that acts on the previous system and modifies its behaviour. Yet, no observable reaction appears. 

The hypothesis of a hidden companion of the particle in QM is not new. The quantum potential of the de Broglie Bohm theory represents an interaction, and therefore there must be another system present. We briefly consider this possibility in the next paragraphs.

\subsection{``Free'' particle, and accompanying system}
%\label{free}

The starting point of the de Broglie Bohm (dBB) theory \cite{Bohm} is well known. If we write the Schroedinger equation of the free spinless electron for the modulus and phase variables we find a continuity equation for the probability density function
$\rho = |\Psi |^2$, and a Hamilton--Jacoby (HJ) like equation for the phase $S$,
$\Psi =  |\Psi | e^{(i S/\hbar ) }$, identified with the action $S$ of usual HJ. This last HJ is that of a free classical particle plus a quantum potential $V_Q$, an odd function of $\rho$ and derivatives. Instead of understanding $V_Q$ as a deterministic interaction as in dBB, we will interpret it as a stochastic term; after all it depends on a probability density. Stochastic interpretations of QM are developed in this or similar ways, e.g. \cite{stoch}. 

We can describe qualitatively some consequences of this quantum interaction. First is the existence of an accompanying system, in interaction with the isolated electron. We can denote it \ae ther or vacuum. Of course, not the classical \ae ther associated to an absolute rest frame. It should be a Lorentz invariant vacuum. There is a Lorentz invariant vacuum in QFT. It is a physical system, playing a fundamental role in the formulation, and it has observable effects, as the Casimir energy. On the other side of length scale there is overwhelming agreement that dark energy, responsible of a large fraction of the cosmic energetic content, is vacuum (zero point) energy. It sounds reasonable, vacuum is dark, it stores energy and there is a lot of vacuum in the universe \cite{S3}.
Perhaps also dark matter could be associated to the vacuum, although it is not the favourite candidate, at least among particle physicist.

 We can not understand in classical terms a Lorentz invariant fluid of mass $m$ particles, with infinite density of particles and energy, as

\[
\Omega = \frac {mc}{p_0} d^3{\bf p} \qquad x^0 = c t 
\]
which is obtained by pulling back, in the momentum space, Minkowsky's  metric into the mass shell $(mc)^2 = p_0^2 - {\bf p}^2$, and then calculating the associated $3$--volume, an explicitly Lorentz invariant definition. Its limit $m\to 0$ obviously vanishes (null vectors in the light cone), but we could, for example, substitute $mc$ by a ``zero point'' momentum $\pi _0$. It could be a classical relativistic approximation of QFT vacuum. 
We have vacuum playing a physical role at cosmological (dark energy) and perhaps galactic (dark matter) scale, and also at the very short length scale of QFT. Vacuum could be the accompanying system of the electron, in the intermediate scale of non relativistic QM; some early proposals about the role of vacuum zero point energy are  \cite{Kal}, \cite{adpo}, \cite{bst}, \cite{st}.  The idea of a real wave in vacuum, different from the complex wave function of QM, can also be found  in the literature \cite{dBro}, \cite{Nel}, \cite{Gros}, \cite{Sbi}.

There are some analogies, just qualitative ones. Vacuum, as a fluid of ``virtual'' particles, interacting with the isolated electron, can be compared to Brownian motion. In Brownian motion there is a kind of uncertainty principle, dispersion on the displacement variable is $\sigma _X = D\sqrt{t}$, so that for the average momentum
$<P(t)> = m(X(t)-0)/t$ (null initial condition) we find $\sigma _{<P>} = m D/\sqrt{t}$, and
$\sigma _X \sigma _{<P>} = m D^2$ constant. 

In Brownian motion we are usually interested in the evolution of the pollen grain, but there is obviously a reaction on the fluid at each collision, giving way to a wavelike motion. A pollen grain in fluid water could be a qualitative image of the wave particle duality, with  interacting systems having properties one of wave and the other of particle. Also the tunnel effect finds here a representation; even if at large time intervals each subsystem preserves energy (statistically), there are fluctuations allowing the pollen grain to overcome a potential barrier. Statistical conservation of energy was proposed very early, in 1924 \cite{bohrks}, although quickly rejected. In the isolated compound particle plus wave system total energy is conserved; yet, interchange of energy can occur between both subsystems. As far as they interact as a whole, these hypothetical internal fluctuations are hidden. 
Hypothetical existence of hidden variables, including a hidden  companion, are the ingredients of the following alternative interpretation of QM.

\section{A local interpretation of Quantum Mechanics}
%\label{local}

In the phase space of a classical point particle, there are individual and real classical paths corresponding to a classical deterministic evolution. In the path integral formalism of QM we consider all virtual paths corresponding to a physical state, and associate  an elementary amplitude
$\Psi _{path} = exp(i S_{path}/\hbar )$ to each of them, with $S_{path}$ the action integral. From it, we calculate the wave function by addition of these amplitudes, {\bf not} probabilities. When we can distinguish  between macroscopically different processes we apply classical probabilities to independent alternatives. There is no understanding, at the classical level, of this mysterious mechanism of the sum of amplitudes. A pure quantum state is described  by a family of virtual paths and amplitudes

\[
{\cal S} = \{ (path _l : \Psi _{path _l})  ; l= 1, \ldots \}
\]
The wave function, in the spatial representation, takes the value

\[
\Psi (q_k ) = \sum _l e^{ \frac {i}{\hbar } S_l(q_k)}
\]
at point $q_k$, where the sum is over all paths with  endpoint $q_k$. The quantum state becomes in this representation

\[
| {\cal S} > = \sum _k \Psi (q_k) |q_k >
\]

\subsection{Alternative interpretation}
%\label{alternative}

The goal of the following  interpretation (as usual) is to describe measurement as an objective process, indistinguishable from other interactions. The alternative interpretation of QM is summarised in the following points

\begin{enumerate}

\item A pure quantum state $\cal S$ (the previous set of virtual paths) is an ensemble of individual states, each individual state with a {\bf label} attached to one of the paths.

\item The labelled path is not a complete description of the individual physical state, there is an accompanying system. No classical probability distribution is assigned to individual paths; instead, the amplitudes of probability encode the statistical information of particle plus wave.  The elementary amplitude $exp(i S_{path}/\hbar )$ is associated to the path. 

\item Outputs are completely determined, at each individual measurement, by the value of the physical magnitude in the labelled path. Relative frequencies are calculated through the usual QM machinery, Born's rule.

\end{enumerate}

In the previous notation, 
\[
{\cal S} = \{ ([path _l, {\cal W}_l] :  \Psi _{path _l}) \}
\]
with ${\cal W}_l$ denoting the accompanying system or wave; one of these $[path _l, {\cal W}_l]$ has a label attached to it in each individual system of the ensemble. As in classical Brownian motion, the wave  depends on the path, not just the end point state of the particle. Each collision modifies the momentum of the particle, and it is also a source of the accompanying wave. Amplitudes of probability encode an ensemble description of particle plus wave, including interference, fluctuations, etc.

The quantum system is observed in our macroscopic world through the label, when an interaction becomes enhanced up to macroscopic size; only a physical magnitude (or family of compatible ones) can be measured at a time because of the unavoidable interaction Hamiltonian properties. So, only partial information of the labelled path is found.

There are inconsistencies between hidden variables and non local self adjoint operators (GHZ). In a QM theory with non local interactions we can not reject the idea of non local measurements. However, if it is the case that interactions are local, then there is obviously no way to perform non local measurements.

\subsection{Hidden states in spatial phase space}
%\label{hidden}

The following recipe could be a way to translate the previous qualitative interpretation of QM from path integral formalism into the Schroedinger representation. We consider a spinless point particle (hidden spin states are presented below), 
with associated Hilbert space ${\cal H}_{QM}$. Definite position state is represented by a vector $|{\bf r}>$, or better by the associated ray in the projective space. It represents the whole set of paths with common end point $\bf r$.

If we want to assign a momentum to the particle at $\bf r$ we need another coordinate. Let us denote  ${\cal H}_{\bf r}$ a Hilbert space corresponding to the vector $|{\bf r}>$, and 
${\cal H}_E = \bigoplus_{\bf r} {\cal H}_{\bf r}$ the direct sum for all points in space. Orthonormal vectors
$|{\bf r},{\bf p}{\bf >}$ generate ${\cal H}_{\bf r}$ (for fixed $\bf r$) and ${\cal H}_E$. In path integral interpretation the physical state corresponding to vector $|{\bf r},{\bf p}{\bf >}$ represents the set of paths with both common end point and final momentum. So, we are just applying a more restrictive grouping of paths. The correspondence is obvious, the set of paths $|{\bf r}>$ is the union of the sets $|{\bf r},{\bf p}{\bf >}$ for all values of $\bf p$. The amplitude associated to ${\bf r}>$ is the sum of amplitudes associated to all $|{\bf r},{\bf p}{\bf >}$. 

The correspondence 

\[
{\cal R} : {\cal H}_E \to {\cal H}_{QM} \quad
{\cal R}(|{\bf r},{\bf p}{\bf >}) = |{\bf r}>
\]
maps vectors in ${\cal H}_{\bf r}$ onto $|{\bf r}>$.

A generic vector in ${\cal H}_E$ is 

\[
|S'{\bf >} = \int d^3{\bf r}d^3{\bf p}\Omega ({\bf r},{\bf p}) |{\bf r},{\bf p}{\bf >}
\]
But this would give way to an heterodox theory, out of QM. Instead, given
$|S> = \int d^3{\bf r}\Psi ({\bf r})|{\bf r}>$ in ${\cal H}_{QM}$ we define

\[
|S'{\bf >} = \int d^3{\bf r}d^3{\bf p} \frac {1}{h^{3/2}} \xi ({\bf p}) 
exp(\frac {i}{\hbar}{\bf p}.{\bf r})|{\bf r},{\bf p}{\bf >} 
\]
such that ${\cal R}(|S'{\bf >}) = |S>$, with $\xi$ the Fourier transform of $\Psi$,

\[
\Psi ({\bf r}) = \frac {1}{h^{3/2}} \int d^3{\bf p} \xi ({\bf p})exp(\frac {i}{\hbar}{\bf p}.{\bf r})
\]
For these particular kind of vectors in ${\cal H}_E$ it is easy to check that operators $X|{\bf r},{\bf p}{\bf >} = x |{\bf r},{\bf p}{\bf >}$ and $P_x|{\bf r},{\bf p}{\bf >} = p_x |{\bf r},{\bf p}{\bf >}$ correspond  to $X \Psi ({\bf r}) = x \Psi ({\bf r})$
and $P_x\Psi ({\bf r}) = -i\hbar \partial _x \Psi ({\bf r})$ in ${\cal H}_{QM}$. 

It must be understood, I insist in this fundamental point, that an individual particle is represented by an in\-di\-vi\-dual state $|{\bf r},{\bf p}{\bf >}$ for the purpose of outputs of measurements, but the whole system particle plus wave is just statistically represented by the orthodox quantum state. This is why we can not associate a classical probability distribution 
$P({\bf r}, {\bf p})$ to the particle state, it is not a free system. Born's rule encodes the statistical description of particle plus wave systems. In particular, it describes interference phenomena, so that the sum of ``individual'' probabilities does not match the compound probability. 

We could also group paths according to its final momentum, $|{\bf p}>$. In this case, the correspondence

\[
{\cal P} : {\cal H}_E \to {\cal H}_{QM} \quad
{\cal P}(|{\bf r},{\bf p}{\bf >}) = exp(-\frac {i}{\hbar}{\bf p}.{\bf r}) |{\bf p}>
\]
maps $|S'{\bf >}$ onto 

\[
{\cal P}(S'{\bf >}) = \,\, \left(\frac {1}{h^{3/2}} \int d^3{\bf r}\right) \,\,\, \int d^3{\bf p}\xi ({\bf p})|{\bf p}>
\]
a divergent expression, but with well defined ray

\[
|S> = \int d^3{\bf r}\Psi ({\bf r})|{\bf r}> =  \int d^3{\bf p}\xi ({\bf p})|{\bf p}>
\]

%%%%%%%%%%%%%%%%%%%%%%%%%%%%%%%%

\subsection{Hidden spin  states}
%\label{hiddspin}

Spin states in QM are described by vectors in a two dimensional Hilbert space ${\cal H}$
generated by, e.g.,  $|+_z>$ and  $|-_z>$.   In a former section  we have considered hidden spin states $|s_1, s_2, \ldots ,$ $s_N{\bf >}$, $s_j = \pm$, determining the output of measurement in an arbitrary direction $\theta _j$ on a plane. A  quantum state is then a linear combination of vectors $|s_1, s_2, \ldots ,$ $s_N{\bf >}$, with complex coefficients $s_1 e^{i\theta _1}$ $+ s_2 e^{i\theta _2}
+ \cdots $, elementary amplitudes $s_je^{i\theta _j}$ playing the role of $exp(\frac {i}{\hbar}S)$. The generalisation from the plane to three dimensional space consist of defining ${\cal H}_{\bf n} $ as a two dimensional vector space over the quaternions, generated by $|+{\bf n}>$ and $|-{\bf n}>$, and assign to the elementary spin state $s {\bf n}$ (spin $s$ in direction $\bf n$), the elementary spin amplitude

\[
\Psi (s {\bf n}) = s ( n_x I + n_y J + n_z K)
\] 
with quaternion numbers $I^2 = -1$, $I J = K$, etc, which have  the algebraic properties of Pauli matrices. 
The total Hilbert space is 

\[
{\cal H}_{SP} = {\cal H}_{{\bf n}_1} \times {\cal H}_{{\bf n}_2} \times \cdots
\]
up to an arbitrary number of directions (lines in the projective space). 
The total amplitude associated to the hidden state $|s_1, s_2,$ $\ldots ,$ $s_N{\bf >}$ is  

\[
\Psi ( s_1 , s_2, \ldots , s_N)
= \sum _{j=1}^n \Psi (s_j {\bf n}_j)
=  (\sum s_j n_{xj}) I  + (\sum s_j n_{yj}) J  + (\sum s_j n_{zj}) K
\]
If we were able to measure simultaneously in two directions, the total amplitude would be

\[
\Psi (s_1, s_2) = \sum_{j \neq 1,2}\,\,\,\, \sum _{s_j = \pm} \Psi ( s_1 {\bf n}_1, s_2 {\bf n}_2, s_3 {\bf n}_3\ldots , s_N {\bf n}_N)
\]
which equals $ \Psi (s_1 {\bf n}_1) + \Psi (s_2 {\bf n}_2) = s_1{\bf n}_1 + s_2{\bf n}_2$ (as quaternions, not vectors). Form this, Born's rule determines the relative frequencies

\[
P(s_1, s_2) = {\cal N} |s_1 {\bf n}_1 + s_2 {\bf n}_2|^2 = 
 {\cal N} 2 ( 1 + s_1 s_2 {\bf n}_1 \cdot {\bf n}_2)
\]
$\cal N$ is the  adequate normalisation, and we have used the identity 

\[
{\bf n}_1^*{\bf n}_2 + {\bf n}_2^*{\bf n}_1 = 2 {\bf n}_1 \cdot {\bf n}_2
\]
where the left hand side is a quaternion expression, and the right hand is vectorial with the usual scalar product.

A  $|+{\bf n}_1>$ spin state of an individual particle in orthodox QM is represented here by the ensemble of all hidden states with $s_1 = +$, 

\[
|+{\bf n}_1> \equiv \{ |s_1 = +, s_2, \ldots , s_N{\bf >}; j>1,  s_j = \pm \}
\]
with corresponding amplitude coefficients; each individual system has a label on some particular hidden state, determining the particular output in arbitrary directions. 
An hypothetical accompanying ``spin wave'' coupled to the particle is statistically represented by the  amplitudes of the ensemble, with the corresponding interference phenomenon. Therefore, no probability distribution for hidden spin states is applied (it does not exist); equivalently, $|s_1, s_2, \ldots ,$ $s_N{\bf >}$ is not a {\bf complete} description of an individual system (particle plus wave), although it completely determines the output of arbitrary spin measurements on the particle.

The orthodox total null spin state of the pair of entangled particles of EPR is represented by the {\bf direct product} of hidden spin states for each individual particle with all outputs with corresponding amplitudes, 

\[
|S=0> \equiv \{ |s^a_1 , s^a_2, \ldots , s^a_N{\bf >};  s^a_j = \pm \} \,\,\times
\]

\[
\{ |s^b_1 , s^b_2, \ldots , s^b_N{\bf >};  s^b_j = \pm \}
\]
such that the labelled  states of each {\bf individual} pair are {\bf correlated} at the generation process,  $(s_1, s_2, \ldots , s_N)_{\lambda}^a$ with $(- s_1, - s_2,$ $\ldots ,$ 
$- s_N)_{\lambda}^b$. Then,  output of an arbitrary individual measurements (if both Alice and Bob select the same direction) is perfectly correlated.
Hypothetical probability distributions for two or more spin directions of an individual particle are calculated as usual through Born's rule, but they can not be observed. In EPR, the measurement performed on $a$ is also an indirect measurement on $b$, and  $P(s_1, s_2)$ becomes observable. But each measurement process is independent in this interpretation, i.e., after $A$ measurement in direction ${\bf n}_1$, particle $a$ output state is $|s^a_1{\bf n}_1>$, but particle $b$ remains in the initial state, with all outputs and corresponding amplitudes. We have just obtained partial  information about the position of the label in the individual $b$ particle state, and this explicit, observed, information must be used when applying Born's rule, as in orthodox QM. In other words, the  conditional probability distribution is different form the initial one, but we apply this conditional probability because of the {\bf information} obtained in $A$ measurement, not because the has been any transmission of physical influence. 

We can compare it with the two slit experiment, where $L$ (or $R$) is the hidden label. If we have information about it, then the $\Psi _L$ (resp. $\Psi _R$) is used in Born's rule. If we do not have this information, we use $\Psi _L + \Psi _R$. A closer analogy between the spin measurement of entangled particles in three directions and a modified two slit experiment  would be to consider a screen with four holes at coordinates $(\pm x_0, \pm y_0)$, and a detector on the final screen able to determine if the electron arrives to one of two regions denoted $+A$ or $-A$. Let us denote $\Psi (s_x,s_y,s_A)$, $s_j = \pm$ the amplitudes for the corresponding paths, e.g., $\Psi(+,+,+)$ represents the path through the hole $(+x_0,+y_0)$ arriving to the region $+A$. 

If we measure 
the position $\pm x_0$ on the first scree and the  position $\pm A$ on the final screen, we get the corresponding table of relative frequencies $P(\pm x_0, \pm A) = |\Psi(\pm,+,\pm) + \Psi(\pm,-,\pm)|^2$. 
In the analogous spin measurement we are measuring particle $a$ in direction ${\bf n}_1$ and particle $b$ in direction ${\bf n}_3$. We can also choose to measure $\pm y_0$ and $\pm A$, respectively directions ${\bf n}_2$ and ${\bf n}_3$, to find $P(\pm y_0, \pm A) = |\Psi(+, \pm,\pm) + \Psi(-,\pm,\pm)|^2$. 

The interference term in $\Psi(\pm,+,\pm) + \Psi(\pm,-,\pm)$ is responsible of the observed correlations, and it can not be reproduced by a classical probability distribution, where either two positive probabilities can not add to zero (two slit experiment), or local hidden variables probability distributions must fulfil Bell's inequalities (entangled spins) and can not reproduce quantum correlations. 

By restricting the analysis to $A$ measurement events in ${\bf n}_1$ direction,  we are determining  two sub--ensembles of particles $b$, those whose label is attached to $b$ spin states with $-s^b_1$ when $s_1^a$ is observed, and  with $s_1^b$ when $-s_1^a$ is observed. Sum of all amplitudes associated to these particle $b$  states and the corresponding $B$ measurement outputs $s_2^b$ in ${\bf n}_2$ direction determines the distribution of probabilities $P(s_1^b, s_2^b) = P(-s_1^a,s_2^b)$, the quantum correlations. In this interpretation the physical phenomenon is local.

\subsection{Summary}
%\label{summary}

Probably the most celebrated alternative interpretation of QM is the dBB Theory, with a quantum potential and deterministic trajectories of point particles. All non hidden variables obstructions find here an explicit counterexample. However, dBB is non local as is QM, and no physical description of the interacting system, responsible of the quantum potential, is proposed.

Wigner's quasi probability distribution in phase space appeared very early but, because of  non positivity of the distribution function, it was rejected. We have seen that the same idea can be applied to spin states.  Quasi probability distributions are the price to pay in order to substitute the mysterious sum of amplitudes rule by a rule closer to classical probabilities, and to introduce a point particle  into the framework. Typical interference phenomena are  linked to the heterodox quasi probability properties, where ``independent'' events can add to a null probability. Neither orthodox QM, nor dBB or quasi probabilities confront the paradox with respect to the action reaction principle.

The proposed alternative interpretation of QM   preserves the quantum machinery for calculating relative frequencies, the mysterious sum of amplitudes and Born's rule, and simply adds a label to the {\bf actual} path in the path integral formalism. This label prescribes the output of measurements in individual systems. An orthodox quantum state is interpreted in this proposal as a statistical ensemble of different physical systems (labelled paths); an individual system is not completely determined by the labelled path, there is a hidden wave like companion.

A mathematical description of hidden spin states, in a Hilbert space over the quaternions, is presented. 
Elementary amplitudes are associated to individual spin states, as in the path integral formalism in space time. The use of amplitudes of probability allows to interpret  EPR  correlations as local phenomena of interference, in complete analogy with the two slit experiment.

The sum of amplitudes rule, taken for granted in orthodox QM and in this alternative interpretation, remains a mystery from the classical point of view. Moreover,  the action reaction principle seems to be violated, at least in indirect measurements. In direct measurement there is not apparent paradox if the system is not  an eigenstate of the physical magnitude, both systems change of state; for eigenstates, if we add a hidden label the paradox can be explained by a hidden reaction (jump of the label) to the action on the pointer. Indirect measurements contradict the action reaction principle unless there is hidden reaction of some kind. 

It is proposed to look for a reaction in the modified  two slit experimental set up with a laser light beam along the $R$ slit. The  hypothetical action (phase shift) on the  accompanying system (vacuum wave) could  generate some reaction on a photon of the beam.

\section{acknowledgements}

Financial support from research project MAT2011-22719 is acknowledged.


\end{document}